\documentclass[aps,prb, amsmath,amssymb,floatfix,12pt]{revtex4-1}
\usepackage{tabularx}
\usepackage{bm}
\usepackage{euscript}
\usepackage{graphicx}
\usepackage{color}
\usepackage{amsfonts}
\usepackage{exscale}
\usepackage{amsbsy}
\usepackage{subfigure}
\usepackage{textcomp}
\usepackage{comment}

\usepackage{hyperref}

\numberwithin{equation}{section}

\newcommand{\be}{\begin{equation}}
\newcommand{\ee}{\end{equation}}
\newcommand{\bea}{\begin{eqnarray}}
\newcommand{\eea}{\end{eqnarray}}

\begin{document}

\title{Metallic quantum critical points with finite BCS couplings}
\author{ S. Raghu$^{{\bar \psi},\psi}$, Gonzalo Torroba$^{\phi}$, Huajia Wang$^{\bar{\psi}}$}
\affiliation{$^{\bar \psi}$Stanford Institute for Theoretical Physics, Stanford University, Stanford, California 94305, USA}
\affiliation{$^\psi$SLAC National Accelerator Laboratory, 2575 Sand Hill Road, Menlo Park, CA 94025, USA}
\affiliation{$^\phi$Centro At\'omico Bariloche and CONICET, Bariloche, Rio Negro R8402AGP,
Argentina}
\date{\today}

\begin{abstract}
We study the fate of superconductivity in the vicinity of a class of metallic quantum critical points obtained by coupling a Fermi surface to a critical boson.  In such systems there is a competition between the enhanced pairing tendency due to the presence of long-range attractive interactions near criticality, and the suppression of superconductivity due to the destruction of the Landau quasiparticles.  We show that there are regimes in which these two effects offset one another, resulting in a novel non-Fermi liquid fixed point with {\it finite}, scale invariant, BCS coupling. While these interactions lead to substantial superconducting fluctuations, they do not drive the system into a superconducting ground state. The metallic quantum critical fixed points are connected to the superconducting regime by a continuous phase transition. These results are established using a controlled expansion in the deviation from $d=3$ spatial dimensions, as well as in a large number $N$ of internal flavors.   We discuss the possible relevance of our findings to the phenomenon of superconducting domes condensing out of a non-Fermi liquid normal state near quantum critical points.  
\end{abstract}

\maketitle

\tableofcontents

\section{Introduction}

One of the central unresolved issues of modern condensed matter physics involves the  enhancement of superconductivity near metallic quantum critical points\cite{Berk1966,Fay1980,Monthoux1994,Abanov2003,Monthoux2007}.  Indeed, many of the strongly correlated electron materials, such as the cuprates\cite{Broun2008}, iron pnictides\cite{Shibauchi2013}, organic\cite{Dressel2011}, and heavy fermion systems\cite{Stewart1984} appear to have enhanced superconducting ``domes" when they are tuned experimentally towards a quantum critical point.  And often, the normal state exhibits scaling behavior that is inconsistent with Fermi liquid theory.  The manner in   which such ``non-Fermi liquid" behavior is related to the  enhanced pairing scale remains a long-standing, yet exciting and actively pursued topic of investigation.

The reason for the enhancement of pairing near quantum critical points has been known for some time: there are induced attractive long-range interactions between  electrons near the Fermi surface mediated by critical order parameter fluctuations. These interactions are long-ranged because of the diverging correlation length at criticality, and, like phonons, order parameter fields mediate  attractive forces. However, precisely the same  order parameter fields  have an opposing effect: they tend to destroy  the quasiparticles, enhancing their scattering rate relative to their energy.  If this second effect is dominant, the effective description of such fermion modes is then no longer governed by Fermi liquid theory.  The fermion fields develop an anomalous dimension, and there is no longer a quasiparticle description of the low energy dynamics.    The system will then be governed by a non-Fermi liquid fixed point.  The destruction of the Landau quasiparticle therefore has a pair-breaking effect, weakening the  superconducting tendency of the system.  

A challenge remains to predict the circumstances under which the enhanced superconducting interaction dominates, and those in which the fermion anomalous dimension dominates. Furthermore, given the fascinating properties of quantum criticality, it would be extremely interesting to construct models of metallic systems exhibiting quantum critical points with non-Fermi liquid behavior in the deep IR. However, in most examples so far, the superconducting instability sets in before non-Fermi liquid effects become important. As a result, the fixed point is fully covered by a superconducting dome, and quantum criticality is not observed.

In this paper we address these questions in a class of quantum metals where the order parameter fields condense at zero momentum.  Examples include the Ising nematic transition in metals, which have been argued to be relevant to the phenomenology of both the cuprate\cite{Kivelson1998} and iron pnictide\cite{Fernandes2014} superconductors.  We consider a solvable large $N$ limit where exact statements about pairing instabilities can be made.  Our first key result is that the competition between the long-ranged attraction and the destruction of Landau quasiparticles  can lead to a fixed point where the BCS interaction among fermions is {\it finite}.  This is in sharp contrast to the behavior of Fermi liquids,  where the BCS coupling flows to zero for repulsive forces, or grows indefinitely, leading to a BCS instability if the couplings are attractive.  The finite BCS interaction fixed point here corresponds to a metallic phase with scale invariant interactions in the BCS channel which do not result in Cooper pair condensation. As a result, we will be able to exhibit a ``naked'' fixed point with critical BCS coupling, not covered by a superconducting dome.\footnote{The fixed point need not extend to zero energy, as there could exist new instabilities at exponentially small scales. However, there will be a parametrically large window of scales where non-Fermi liquid effects dominate, and quantum criticality is observed.} The thermodynamic and transport signatures at such finite BCS interaction fixed points are interesting in their own right and may be relevant to experiments involving quantum critical metals.  We will study these phenomenological properties in future work. 

Our second main result is that there is a continuous transition between the regime where the RG flows are always towards enhanced superconductivity, and the regime where IR stable fixed points with finite BCS couplings occur. Approaching the transition from the critical regime, the IR fixed point annihilates against an unstable UV fixed point and disappears. From the other side, the superconducting parameter and all its derivatives vanish as we tune towards the transition. This is reminiscent of the  Berezinski Kosterlitz Thouless (BKT) transition, as  we will explain in detail below.  

The paper is organized as follows.  In section \ref{sec:sd}, we define the bare tree-level action and, neglecting superconductivity for the moment, we  describe the dominant quantum corrections by solving the Schwinger-Dyson (SD) equations for the system.  These equations are exact to all orders in perturbation theory, for the large $N$ limit we consider. Next, in section \ref{sec:RG} we study the system using the renormalization group. We construct a scaling theory consistent with the analysis of the SD equations, determine the one loop beta functions, and characterize the non-Fermi liquid fixed point. Section \ref{sec:SC} is devoted to the analysis of the BCS interaction: we show that there is a regime where the 4-Fermi coupling flows to a stable fixed point, and another regime where it leads to a superconducting instability. We establish that both states are connected by a continuous phase transition. Section \ref{sec:discussion} contains our conclusions regarding the phase diagram of the theory and discusses future directions. Some technical calculations are presented in two appendices.

%%%%%%%%%%%%%%%%%%%%%%%%%
%%%%%%%%%%%%%%%%%%%%%%%%%
%%%%%%%%%%%%%%%%%%%%%%%%%
%%%%%%%%%%%%%%%%%%%%%%%%%
\section{Effective action and quantum corrections}\label{sec:sd}

In this section we present the classical theory and compute quantum effects. We use the Schwinger-Dyson equations for the boson and fermion two point functions and work in a large $N$ limit, which will allow us to obtain results that are exact to all orders in perturbation theory. Here we focus on the correlation functions that are local on the Fermi surface, while in sections \ref{sec:RG} and \ref{sec:SC} we take into account the 4-Fermi interaction in the BCS channel.

%%%%%%%%%%%%%%%%%%%%%%%%%
%%%%%%%%%%%%%%%%%%%%%%%%%
\subsection{The model}\label{subsec:model}

In this work we will analyze the quantum theory for a Fermi surface coupled to a gapless boson $\phi$.
Our starting bare euclidean Lagrangian  is 
\begin{eqnarray}
L &=& \frac{1}{2} {\rm Tr} \left[ \left( \partial_{\tau} \phi_0 \right)^2 + \left( \nabla \phi_0 \right)^2 \right] + \psi^{\dagger}_{i 0} \left( \partial_{\tau} + \varepsilon_0(i \nabla) - \mu_F \right) \psi^i_0 + L_{\psi, \phi} + L_{BCS} \nonumber \\
L_{\psi, \phi} &=&  \frac{g_0}{\sqrt{N}} \phi^{i}_{0 j}(q) \psi^{\dagger}_{0 i}(k+q) \psi_0^j (k) \nonumber \\
L_{BCS} &=& - \frac{v_0}{2 k_F^{d-1}} \frac{\lambda_0}{N} \psi^{\dagger}_{0 i}(p+q) \psi_0^j(p)  \psi^{\dagger}_{0 j}(-p-q) \psi^{i}_0(-p)\,.
\end{eqnarray}
The subscript `$0$' denotes bare quantities (we will consider the effects of renormalization after integrating out high energy modes below).  The bare band dispersion of the fermions is denoted by $\varepsilon_0(k)$ and for simplicity we consider a rotationally invariant Fermi surface; the chemical potential is $\mu_F = \varepsilon_0(k_F)$.  The sign in $L_{BCS}$ is such that $\lambda_0>0$ corresponds to an attractive interaction.

We will consider a soluble  limit of the theory above.  First, we introduce an internal $SU(N)$ global flavor symmetry (a generalization of spin rotation symmetry) under which the fermion fields transform in the fundamental (vector) representation, whereas the bosons transform in the adjoint (matrix) representation. We work in the limit $N \gg 1$ with $g_0$ and $\lambda_0$ fixed; many diagrams will be shown to be subleading, and it will be possible to resum exactly the leading quantum corrections. We note that the large $N$ theory here explores a distinct asymptotic regime  than  the standard large $N$ approach to this problem, in which the boson remains a scalar while the fermions are fundamental fields of a global flavor symmetry group. Furthermore, we work in $d=3-\epsilon$ spatial dimensions with $\epsilon \ll 1$, namely near the critical dimension for the Yukawa coupling. As discussed below, the small parameter $\epsilon$ will be used to avoid infrared divergences from corrections to the cubic vertex.

Besides providing limits where quantum corrections simplify, $N$ and $\epsilon$ will also affect the infrared dynamics of the theory.
Our task will be to determine the low energy phase diagram of the theory as a function of $N$ and $\epsilon$. Before proceeding to the discussion of quantum corrections, let us develop some intuition by comparing the scales of non-Fermi liquid effects and superconductivity. Anomalous dimension corrections become important at a scale
\be
\mu_{NFL} \sim e^{-\frac{1}{2\gamma(\Lambda)}}\Lambda
\ee
where
\be
\gamma=\frac{g^2}{24\pi^2v} \equiv \frac{\alpha}{2}\,,
\ee
and $g$ and $v$ are physical couplings (see below). We will deduce this result shortly, but for now we just want to explore some of its consequences. On the other hand, the scale of the superconducting gap taking into account the enhancement from boson exchange is (see [\onlinecite{Son:1998uk}] and below)
\be
\mu_\text{sc} \sim e^{-\frac{\pi}{2}\sqrt{\frac{N}{\alpha(\Lambda)}}}\Lambda\,.
\ee
At the quantum critical point described below, $\alpha\sim \epsilon$.
We see here the interplay between NFL and gap effects: for $\epsilon \ll 1/N$, superconductivity dominates, and the Fermi surface is gapped before the NFL regime is reached. However, in the opposite limit $N \gg 1/\epsilon$, $\mu_{NFL} \gg \mu_\text{sc}$ and hence we expect (and will find) strong NFL corrections to the superconducting gap. This is the range where a new quantum critical point for the BCS interaction will obtain. We will show that both regimes are connected by a continuous phase transition that occurs when $N \epsilon \sim 1$.

Our strategy will be to first determine the dynamics in the non-Fermi liquid regime, corresponding to $N \gg 1/\epsilon$. Here we will neglect the superconducting gap, and then check that this is a self-consistent approximation. We will then incorporate effects from superconductivity and will characterize this phase that occurs when $1 \ll N \ll 1/\epsilon$. This will be done by analyzing the RG $\beta$ function for the 4-Fermi BCS coupling.

%%%%%%%%%%%%%%%%%%%%%%%%%
%%%%%%%%%%%%%%%%%%%%%%%%%
\subsection{Quantum corrections}\label{subsec:quantumcorr}

Let us then begin the analysis of quantum effects focusing on the self-energies and boson-fermion coupling.
At sufficiently high energies, the scaling behavior of the theory can easily be understood.  Considering the Yukawa coupling to be small, scaling is constructed about the  limit wherein a Fermi liquid is decoupled from the order parameter field.  Scaling behaviors of the fermions and bosons are governed by Fermi liquid theory and Landau-Ginzburg-Wilson effective theories respectively.  The dimension of any composite operator, such as $\psi^{\dagger} \psi \phi$ is then immediately known from the decoupled scaling dimensions of these fields.  The conclusion of such an analysis is that the Yukawa coupling has a bare scaling dimension 
\begin{equation}\label{eq:classicg}
\left[ g \right] = \frac{3-d}{2} = \frac{\epsilon}{2}\,.
\end{equation}
Our next task is to determine the way in which quantum corrections alter this behavior. 

As discussed in many previous works\cite{Hertz1976, Millis1993, Altshuler1994, Polchinski1994, Lee2009, Yamamoto2010, Mross2010, Mahajan2013,Fitzpatrickone, FKKRtwo, Torroba:2014gqa, Fitzpatrick:2014cfa, Lee2013}, there are many subtleties involved with taking a strict large $N$ limit in this class of theories.  Indeed, the scaling behavior in the large $N$ limit depends quite strongly on the order in which the $N \rightarrow \infty$ and $\omega \rightarrow 0$ limits are taken.  If the $N\rightarrow \infty$ limit is taken first, the resulting fixed points obtained only govern behavior at intermediate energy scales \footnote{to be precise, the description breaks down below the scale $M_D$ to be defined below}, because they only take into account a subset of the important quantum corrections.  In particular, in this limit,  key $\mathcal{O}(1/N)$ quantum corrections that qualitatively alter the IR behavior are neglected; these quantum corrections act as effective relevant coupling constants that destabilize potential $N=\infty$ fixed points.  Here, we wish to avoid such peculiarities, and build in all the important quantum corrections, even those that formally are $1/N$ corrections,  into our theory.  

\begin{figure}[h!]
\centering
\includegraphics[width=0.9\textwidth]{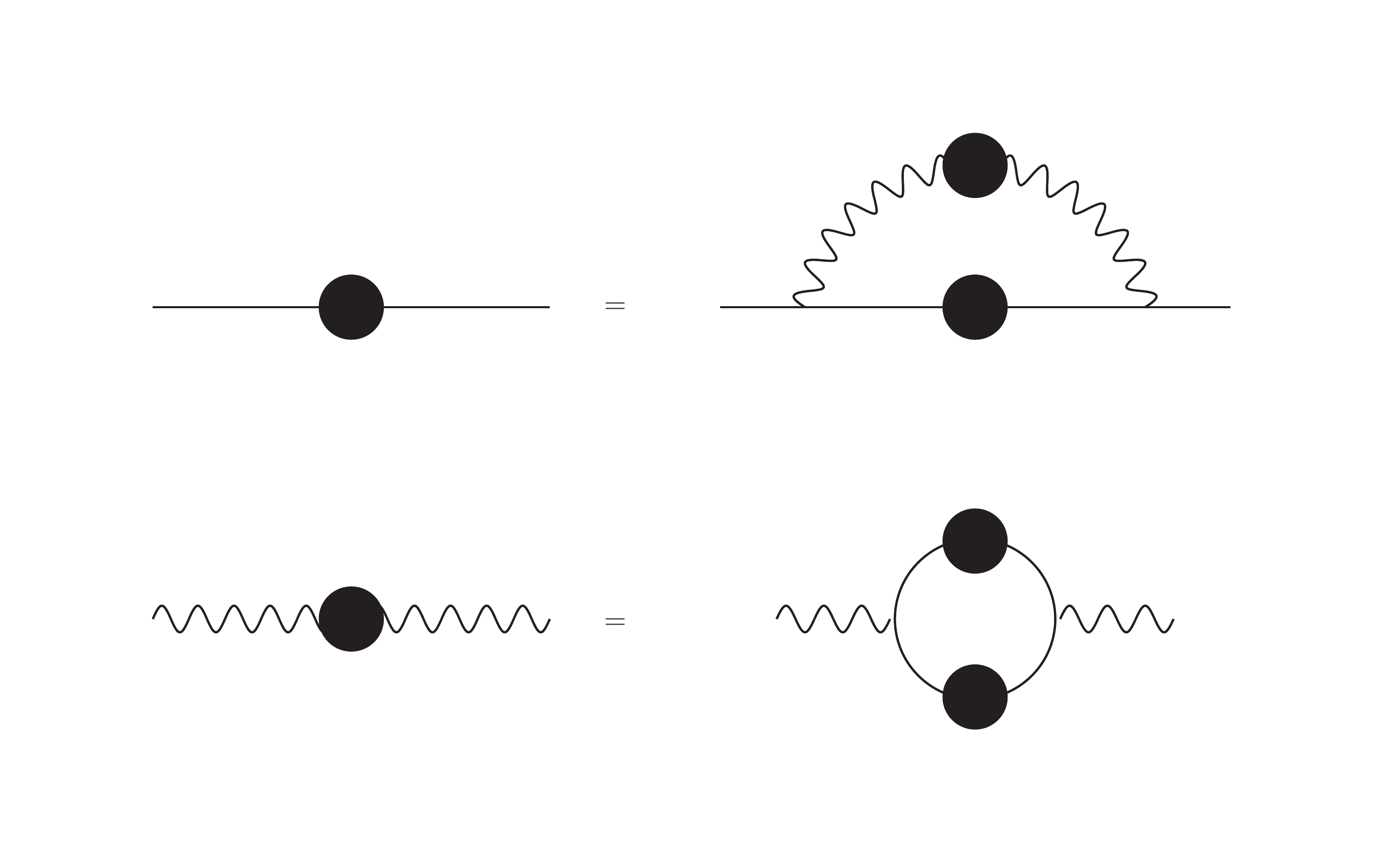}
\caption{\small{Schwinger-Dyson equations for the fermion (full line) and boson (wiggly line) two-point functions in the large $N$ theory.}}\label{fig:SDeq}
\end{figure}

Our strategy for obtaining a scaling theory will be to look for effects that are exact to all orders in perturbation theory at large $N$. This will be obtained by investigating the Schwinger-Dyson (SD) equations for this system. A key simplification of large $N$ is that quantum corrections to the cubic vertex are suppressed by an extra power of $1/N$ compared to the tree level term, at fixed $g_0$. This allows us to neglect vertex corrections.\footnote{Small $\epsilon$ here is also important, so that the infrared divergences found in \cite{Lee2009} do not appear.} In this case we find a closed system of SD equations for the boson and fermion self-energies, expressed as follows (see
Fig. \ref{fig:SDeq}):
\begin{eqnarray}
\label{SD}
\Pi(q_0, q) &\equiv& D^{-1}(q_0, q) - D_0^{-1}(q_0, q) = \frac{g_0^2}{N}\int \frac{d k_0 d^d k }{(2 \pi)^{d+1}}\, G(k_0, k) G(k_0 + q_0, k+q) \nonumber \\
\Sigma(p_0, p) &\equiv& G^{-1}(p_0, p) - G_0^{-1}(p_0, p) = - g_0^2 \int \frac{d k_0 d^d k }{(2 \pi)^{d+1}}\, G(k_0, k) D(p_0 - k_0, p-k)\qquad
\end{eqnarray}
where $D(G)$ refer to the exact boson (fermion) propagator.  

There are many possible solutions to these equations, depending on the parameter that is being held fixed while taking the large $N$ limit.  We will solve these equations, in $d=3-\epsilon$ spatial dimensions, and in the limit where both $N\rightarrow \infty$, $k_F \rightarrow \infty$ holding fixed the following quantity
\begin{equation}
M_D^2 = c_d\, \frac{k_F^{d-1}}{ 2\pi v_0}\, \frac{g_0^2}{N},
\end{equation}
where $c_d$ is a constant that depends on the dimensionality of space.  The factor $k_F^{d-1}/v_0$ is proportional to the density of states at the Fermi energy.  
Physically, this is the scale below which Landau damping of the boson becomes very important and is equivalent to the ``Debye'' screening scale, below which long range Coulomb interactions are screened in a metal.  
As we will see below, by holding $M_D^2$ fixed, the theory simplifies substantially in the IR.  
These equations build in the dominant quantum corrections in the large $N$ limit above the scale of the superconducting gap.  

Above the scale of $M_D$, there are logarithmic corrections to the fermion self-energy, which are both frequency and momentum-dependent.   They produce a  small anomalous dimension (proportional to $\epsilon$) and cause a slight reduction of the Fermi velocity.  Both of these effects can be seen directly in perturbation theory in the UV.  For large enough $M_D$, however, these effects are subdominant in relation to the Landau damping of the bosonic order parameter fields, which is a UV finite, non-local quantum correction to  $D(q)$, and therefore is invisible in a Wilsonian treatment of the problem.  Nevertheless, it strongly affects the dynamic scaling of  the boson.  
This quantum correction  comes directly from the first equation in (\ref{SD}) and can be interpreted as coming from resumming the geometric series of fermion bubbles (analogous to RPA):
\be
D(q_0, q)^{-1}=q_0^2+\vec q^{\,2}+\Pi(q_0, q)\,.
\ee
The behavior in the regime $|q_0| \ll v_0 |\vec q\,|$, which will be relevant for us, is\cite{Torroba:2014gqa}
\be
\Pi(q_0, q) \approx M_D^2\,\frac{|q_0|}{v|\vec q\,|}\;\;.
\ee
 The boson propagator then takes the approximate form 
\be\label{eq:Dz3}
D(q_0, q)^{-1}\approx\vec q^{\,2}+M_D^2\,\frac{|q_0|}{v_0|\vec q\,|}
\ee
and is characterized by a $z_b=3$ scaling $q_0 \sim |\vec q\,|^3/M_D^2$.  $M_D$ therefore acts as a crossover scale, separating the $z=1$ UV behavior from the $z=3$ IR behavior in this regime.  Our analysis will focus on scales smaller than $M_D$.

Below the scale $M_D$, the self-energy ceases to have substantial momentum dependence and depends mainly on frequency.   Therefore, at energies much less than $M_D$ the velocity renormalization is primarily due to the fermion anomalous dimension (i.e. the velocity will vanish at the fixed point where $Z$ vanishes).   The non-zero anomalous dimension  will also cause the renormalization of the Yukawa coupling (note again that vertex corrections are suppressed in our large $N$ limit; only field rescaling due to the anomalous dimension causes renormalization of the bare Yukawa coupling).  Given (\ref{eq:Dz3}), the solution to the fermion SD equation in (\ref{SD}) is
\be
\Sigma(p_0) = -i p_0\,\frac{3 \alpha_0}{\epsilon}\left( M_D^2 \vert p_0 \vert \right)^{-\epsilon/3}\,,\qquad \alpha_0 \equiv \frac{g_0^2}{12 \pi^2 v_0} \,.
\ee
Recalling that the bare coupling $\alpha_0$ has engineering dimension $\epsilon$, the self-energy has engineering dimension $1$ (as it should), but the scaling dimension with $p_0$ is $1-\epsilon/3$. We emphasize again that this scaling applies below $M_D$, which plays the role of a UV scale in our effective theory.

We conclude that the effective (quantum) Lagrangian involving the fermion kinetic energy and the Yukawa coupling is 
\begin{equation}
L_{eff, \psi} = -\psi_0^{\dagger} \left[ i Z(p_0)p_0 - v_0 p_{\perp} \right] \psi_0 + \frac{g_0}{\sqrt{N}} \psi^{\dagger}_0(k+q) \psi_0(k) \phi_{0 } (q)
\end{equation}  
where $Z(p_0)$ is the quasiparticle residue,
\be\label{eq:Z1}
Z(p_0) =1- \frac{\Sigma(p_0)}{ip_0}= 1 + \frac{3 \alpha_0}{\epsilon}\left( M_D^2 \vert p_0 \vert \right)^{-\epsilon/3}\,.
\ee
The fermion momentum here is decomposed radially towards the Fermi surface,
\be\label{eq:radialp}
\vec p = \hat n (k_F+ p_\perp)\,,
\ee
and $\hat n$ is a unit vector that defines the position on the Fermi surface.

The second term in (\ref{eq:Z1}) represents the effects of quantum corrections.  Below  a scale $\mu_{NFL}$ defined by 
\begin{equation}
\mu_{NFL} = \left( \frac{3 \alpha_0}{\epsilon} \right)^{3/\epsilon} M_D^{-2}\,, 
\end{equation}
the quasiparticle residue $Z$ is dominated by quantum corrections, and the frequency dependence of the fermion kinetic term has the following behavior:
\begin{equation}
p_o  \ll \mu_{NFL}: \ \ Z(p_0) p_0 \approx  \frac{3 \alpha_0}{\epsilon}  M_D^{-2\epsilon/3}\,p_0^{1-\epsilon/3}\,.
\end{equation}
This is of the form $p_0^{1-2 \gamma} $: in  other words, at low energies the fermions develop an anomalous dimension, 
\begin{equation}
2 \gamma \approx \frac{\epsilon}{3}
\end{equation}
and the quasiparticle residue becomes
\be\label{eq:ZIR}
Z(\mu) \approx  \left(\frac{\mu_{NFL}}{\mu} \right)^{\epsilon/3}\,.
\ee

%%%%%%%%%%%%%%%%%%%%%%%%%
%%%%%%%%%%%%%%%%%%%%%%%%%
\subsection{Physical quantities and IR dynamics}\label{subsec:IR}

The bare quantities written above are not physically measurable.  Once quantum corrections arise, the physically observable fields and couplings depend on the energy scale.  In the large $N$ limit this occurs only because of 1) Landau damping of the bosons and 2) a non-zero fermion anomalous dimension.  

The physical quantities are obtained after canonically normalizing the fermion fields at a scale $\mu$, $\psi \equiv Z^{1/2}(\mu) \psi_0$. The physically observable Fermi velocity (as seen in heat capacity or in tunneling density of states, for instance) will depend on the energy scale $\mu$ as
\begin{equation}
\label{eq:vz}
 v(\mu) = Z^{-1}(\mu) v_0 \,.
\end{equation}
Note that for $\mu \ll \mu_{NFL}$, the velocity flows to zero with a power-law determined by (\ref{eq:ZIR}),
\be
v(\mu) \sim v_0 \left(\frac{\mu}{\mu_{NFL}}\right)^{\epsilon/3}\,.
\ee 
Next, we have to determine the relation between the physical dimensionless coupling $g(\mu)$ and $g_0$ or, more usefully,\footnote{Indeed, a field redefinition shows that in the $z_b=3$ regime, the perturbative expansion is given in terms of $\alpha(\mu)$.}
\be\label{eq:alphadef}
\alpha(\mu) \equiv \frac{g^2(\mu)}{12 \pi^2 v(\mu)}\,.
\ee
From the effective Lagrangian in terms of canonical fields, we read off $\alpha(\mu) \propto Z(\mu)^{-1} \alpha_0$. Here two powers of $Z(\mu)^{-1}$ come from $g^2(\mu)$ in (\ref{eq:alphadef}), while $v(\mu)$ gives an additional factor of $Z(\mu)$ according to (\ref{eq:vz}). 

It remains to determine the dimension $\delta$ of $\alpha(\mu)$. For this, we write the relation between bare and renormalized couplings as
\be\label{eq:alpha2}
 \alpha_0=M_D^{\epsilon-\delta}\,Z(\mu)\, \mu^\delta\,\alpha(\mu) \,,
\ee
where $M_D$ is added to match engineering dimensions. The dimension $\delta$ by definition cancels factors of the external frequency in loop integrals, such that we get a perturbative expansion in terms of $\alpha$ and dimensionless ratios $\mu/p_0$. The one loop fermion self-energy calculated using renormalized perturbation theory is
\be
\Sigma(p_0)= - i p_0\,\frac{3 \alpha(\mu)}{\epsilon}\,M_D^{\epsilon/3-\delta}\frac{\mu^\delta}{|p_0|^{\epsilon/3}}\,.
\ee 
Therefore, $\delta=\epsilon/3$, namely
\be\label{eq:SDdim}
[\alpha(\mu)]=\frac{\epsilon}{3}
\ee
and, as expected, the microscopic scale $M_D$ cancels when working in terms of the physical coupling.

Combining this with (\ref{eq:alpha2}) obtains the relation between $\alpha_0$ and $\alpha(\mu)$ to all-orders in the large $N$ theory,
\be\label{eq:alphamu}
\alpha(\mu)= \frac{\alpha_0}{\frac{3\alpha_0}{\epsilon}+\left(M_D^2\mu \right)^{\epsilon/3}}\,.
\ee
This energy dependence of $\alpha(\mu)$ has the property that for scales $\mu \ll \mu_{NFL}$ the coupling flows to
\be
\alpha(\mu) \to \frac{\epsilon}{3}\,,
\ee
a non-Fermi liquid fixed point that will be analyzed in more detail in Sec. \ref{sec:RG}.

Eqs. (\ref{eq:vz}) and (\ref{eq:alphamu}) are the main results of this section. They tell us how the physical fermion velocity and Yukawa coupling at a scale $\mu$ behaves. 
Thus, while perturbation theory about the decoupled fermion-boson limit would suggest that $\left[ \alpha \right] = \epsilon$, quantum corrections treated by the SD equations instruct us how these classical dimensions are altered.  
 
To summarize: using the SD equations we have computed the quantum corrections that occur to all orders in perturbation theory in the large $N$ limit, at scales $E<M_D$, and with $M_D$ fixed. Our next objective is to obtain this behavior using scaling and Wilsonian RG below the Debye scale. This method will then allow us to determine the evolution of the BCS interaction. An alternative analysis of the superconducting gap in terms of its Schwinger-Dyson equation (the Eliashberg equation) will be presented in [\onlinecite{RTWlong}].

%%%%%%%%%%%%%%%%%%%%%%%%%
%%%%%%%%%%%%%%%%%%%%%%%%%
%%%%%%%%%%%%%%%%%%%%%%%%%
%%%%%%%%%%%%%%%%%%%%%%%%%
\section{Renormalization group approach including BCS couplings}\label{sec:RG}

We next determine the RG $\beta$ functions that define the flow of physical couplings for our theory. The first step will be to reproduce the SD results in a Wilsonian RG framework. We will then focus on the RG flow for the BCS coupling. A Schwinger-Dyson treatment of superconductivity effects including the anomalous dimension is more involved, and will be presented in~\cite{RTWlong}. Before proceeding, we note that in the theory in $d=3-\epsilon$ spatial dimensions, it will be computationally more convenient to organize quantum corrections in powers of $1/\epsilon$ (effectively using $\epsilon$ as a regulator), instead of employing the physical cutoffs $\Lambda_f$ and $\Lambda_b$. The map between both approaches was given in~\cite{Fitzpatrick:2014cfa}.

%%%%%%%%%%%%%%%%%%%%%%%%%
%%%%%%%%%%%%%%%%%%%%%%%%%
\subsection{Scaling theory}\label{subsec:scaling}

The first step in an RG approach is to construct a consistent scaling.
The solution of the SD equations places strong constraints on a scaling theory which, in particular, has to reproduce (\ref{eq:SDdim}). We now present the scaling that agrees precisely with the form of quantum corrections obtained above.  

Ref.~[\onlinecite{Fitzpatrick:2014cfa}] showed that a consistent renormalization of the Fermi surface coupled to a massless boson requires two independent decimation procedures: the Fermi surface high momentum modes are integrated on shells
\be
\Lambda_f-d \Lambda_f< |p_\perp|< \Lambda_f\,,
\ee
where the fermion momentum is decomposed radially towards the Fermi surface,
\be
\vec p = \hat n (k_F+ p_\perp)\,,
\ee
and $\hat n$ is a unit vector that defines the position on the Fermi surface.
On the other hand, the boson momentum $\vec q$ is decimated towards the origin, with an independent cutoff $\Lambda_b$:
\be
\Lambda_b-d \Lambda_b< |\vec q\,|< \Lambda_b\,.
\ee
These two independent momentum-shell integrations are needed to capture the leading quantum corrections to correlation functions. The reason is that some contributions that look IR from the point of view of the fermions, actually come from UV bosonic modes, and hence have to be taken into account in the Wilsonian RG. Important consequences of this were the running Fermi velocity, and tree-level logarithmic running of 4-Fermi couplings.

This approach has to be modified if the boson has a nontrivial dynamical exponent $z_b$. This is discussed in detail in Appendix \ref{app:scaling}. First, we find that the scaling of the fermions is not changed by $z_b$,
\be
[p_0]= [p_\perp]=1\;,\;[\psi(p)]=-3/2\,.
\ee
The dimension of the 4-Fermi BCS coupling is classically marginal in any dimension, since we work in the spherical RG for the fermions.

On the other hand, the bosonic scaling is modified as follows. Given a patch with angular position $\hat n$ on the Fermi surface, we decompose the boson momentum into orthogonal components
\be
\vec q= \hat n q_\perp + \vec q_\parallel\,.
\ee
We show in the Appendix that the correct scaling obeys
\be\label{eq:boson-scaling1}
[q_0]=[q_\perp]=1\;,\;[q_\parallel]=1/3\;,\;[\phi(q)]=-\frac{10-\epsilon}{6}
\ee
for $d=3-\epsilon$. 

With these scalings, the classical dimension of $g$ becomes
\be\label{eq:gscaling}
[g]=\frac{\epsilon}{6}\,.
\ee
Therefore, the dimension of $g$ is nearly marginal with this scaling for the overdamped boson, as is also the case above the Landau damping scale.  This scaling reproduces the quantum result (\ref{eq:SDdim}) obtained from solving the SD equations.

The near-marginality of $g$ after including Landau damping is important, as it is consistent with a smooth crossover between the undamped and overdamped regimes (see also~\cite{Torroba:2014gqa}), and perturbation theory does not break down. Had we scaled the boson momentum homogeneously, the result would have been $[g]=(2+\epsilon)/6\approx 1/3$, giving an order one relevant interaction in the overdamped regime. Such a relevant coupling would be inconsistent with the results of the SD equations.

%%%%%%%%%%%%%%%%%%%%%%%%%
%%%%%%%%%%%%%%%%%%%%%%%%%
\subsection{Local non-Fermi liquid behavior}\label{subsec:localNFL}

\begin{figure}[h!]
\centering
\includegraphics[width=1.1\textwidth]{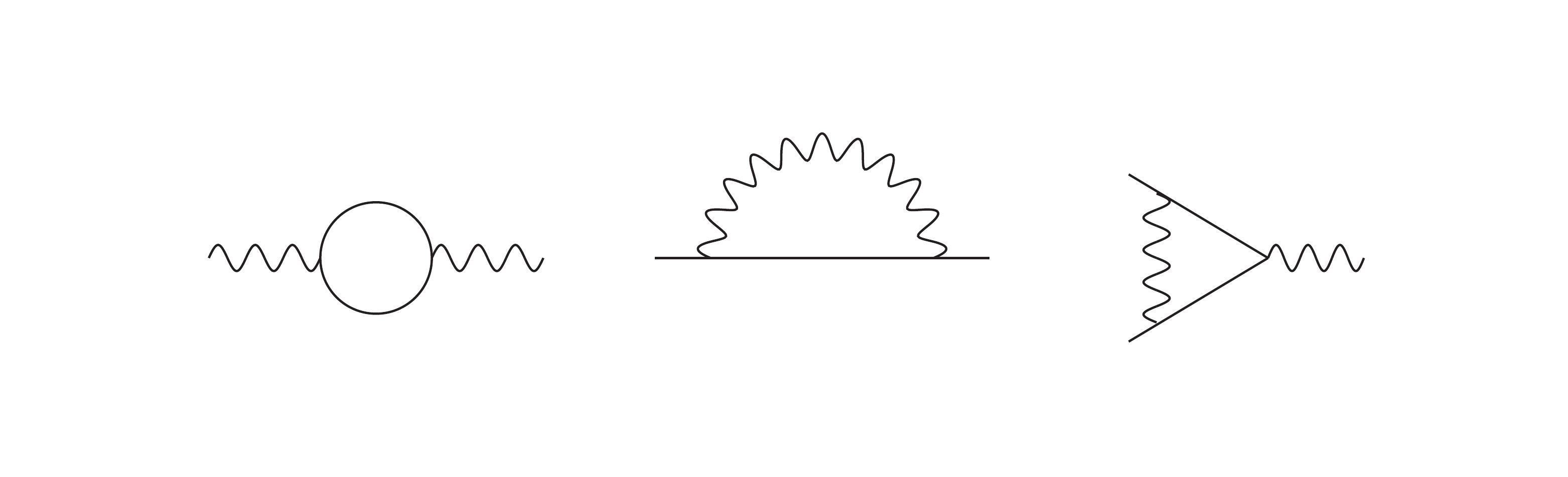}
\caption{\small{One loop corrections to the boson self-energy, fermion self-energy and cubic Yukawa vertex.}}\label{fig:local-correlators}
\end{figure}

Let us discuss first the RG for correlations that are local on the Fermi surface --the self-energies and the boson-fermion coupling. The one loop corrections are shown in Fig.~\ref{fig:local-correlators}.

We argued in Sec.~\ref{sec:sd} that below the Landau damping scale the boson flows to a $z_b=3$ dynamical exponent. This is not a Wilsonian effect: it comes from integrating particle-hole pairs at the Fermi surface; it is a finite renormalization effect that becomes relevant below the scale $M_D$. Here we start from this dressed boson propagator and focus on $E<M_D$, but we note that the crossover between $z_b=1$ and $z_b=3$ is smooth\cite{Torroba:2014gqa}.

In the $z_b=3$ regime, the fermion self-energy depends only on frequency. This results in a velocity flowing to zero with a rate determined by the anomalous dimension, $Z_v=Z^{-1}$. Furthermore, quantum corrections to the cubic vertex are suppressed by $1/N$ at large $N$, and can be neglected in our perturbative framework. Therefore, the one loop $\beta$ functions characterizing quadratic and cubic correlators are determined purely in terms of the fermion anomalous dimension $\gamma$. Using renormalized perturbation theory at one loop, the anomalous dimension is~\cite{Torroba:2014gqa}
\be
\gamma=\frac{g^2}{24\pi^2v}\,.
\ee
In terms of $\alpha$ defined in (\ref{eq:alphadef}), the one loop $\beta$ functions on a local patch of the Fermi surface then become
\bea\label{eq:betafcs1}
&&2\gamma= -\mu\frac{d\log Z}{d\mu}=\alpha\nonumber\\
&&\mu\frac{d \log v}{d\mu} = 2\gamma  \\
&&\mu\frac{d\alpha}{d\mu} =-\frac{\epsilon}{3} \alpha + 2\gamma \alpha\,. \nonumber
\eea
This agrees with the $\beta$ functions obtained from the SD analysis by requiring that the bare parameter in the relation (\ref{eq:alphamu}) be independent of the RG scale $\mu$. Note that the one loop approximation here is exact at large $N$.

This system admits a non-Fermi liquid fixed point,
\be
\alpha_*=\frac{\epsilon}{3}\;,\;\gamma_*=\frac{\epsilon}{6}
\ee
that is perturbatively controlled at small $\epsilon$ and large $N$. Although we can make no definite prediction for $\epsilon \sim 1$, we note that as $\epsilon \to 1$ (i.e. for $d=2$ spatial dimensions) this fixed point approaches the strongly coupled non-Fermi liquid of~\cite{Polchinski1994}. It would be interesting to use the $\epsilon$ expansion to understand more systematically the $\epsilon \to 1$ limit.

%%%%%%%%%%%%%%%%%%%%%%%%%
%%%%%%%%%%%%%%%%%%%%%%%%%
\subsection{BCS $\beta$ function}\label{subsec:BCSbeta}

\begin{figure}[h!]
\centering
\includegraphics[width=1.1\textwidth]{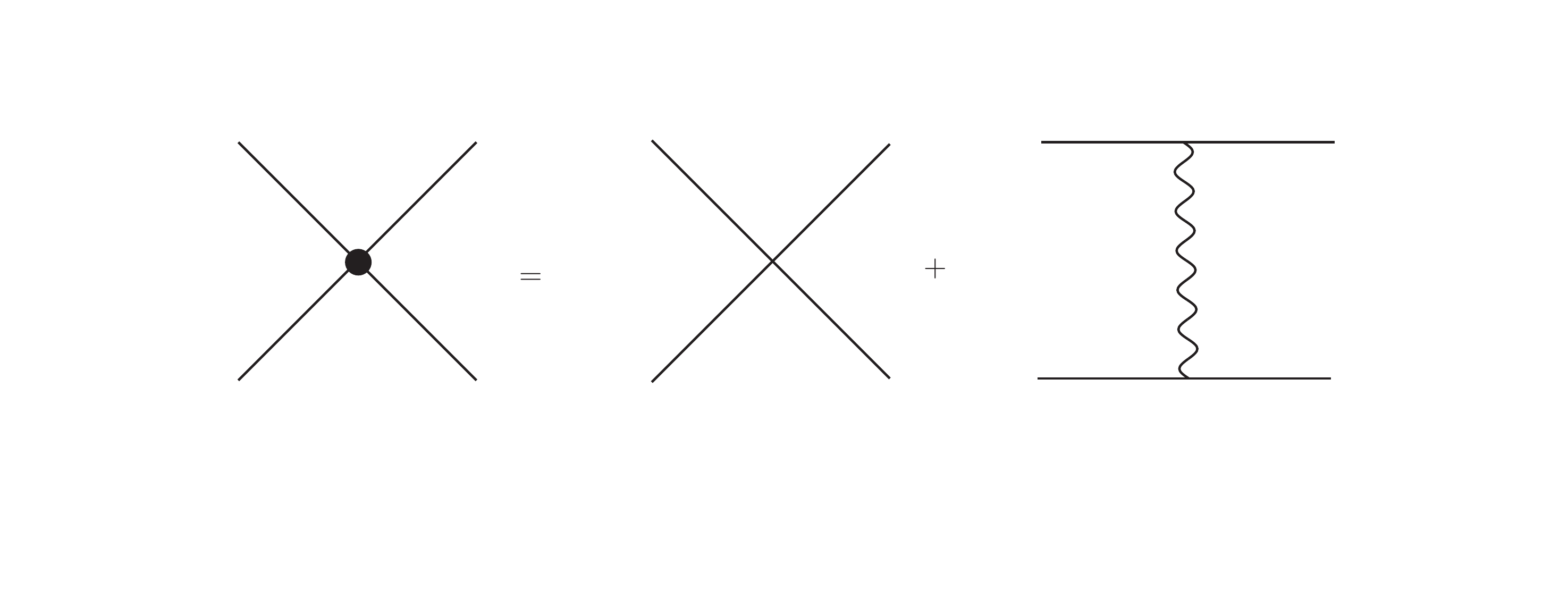}
\caption{\small{Tree-level running of the BCS interaction.}}\label{fig:4Fermi}
\end{figure}

\begin{figure}[h!]
\centering
\includegraphics[width=1.1\textwidth]{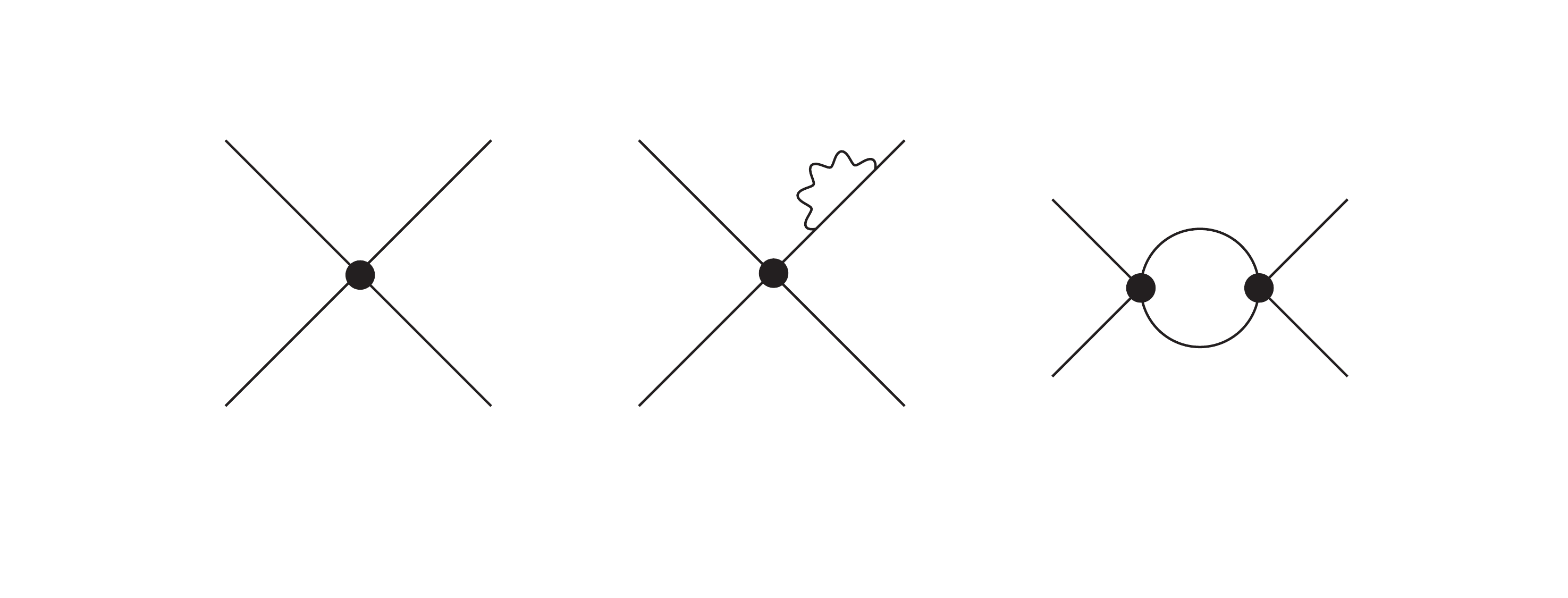}
\caption{\small{Diagrammatic one loop contributions to the BCS coupling.}}\label{fig:4Fermiloop}
\end{figure}

We now want to include the effects from the BCS 4-Fermi coupling, which is classically marginal and can destabilize this fixed point. The renormalization of the BCS interaction proceeds in two steps\cite{Fitzpatrick:2014cfa, Fitzpatrick:2014efa}. First, at tree level the boson exchange gives rise to running of the BCS couplings in the angular momentum basis; see Fig.~\ref{fig:4Fermi}. At one loop there are two additional contributions: from the anomalous dimension and the BCS one loop diagram. This is shown in Fig.~\ref{fig:4Fermiloop}.

The one loop beta function then becomes (see also\cite{Son:1998uk})
\be\label{eq:betaBCS}
\mu\frac{d\lambda}{d\mu} = -2 \pi^2 \alpha+2\gamma \lambda - \frac{\lambda^2}{2\pi^2 N}\,.
\ee
(We recall that $\lambda>0$ corresponds to an attractive interaction.) 
To our knowledge, the term linear in $\lambda$ has not been included in previous works. It captures the non-Fermi liquid corrections to fermion scattering on antipodal points of the Fermi surface and hence the formation and condensation of Cooper pairs. We will find that it has dramatic effects on the IR phase structure of the theory, which we consider in Sec.~\ref{sec:SC}.

There are two and higher loop quantum corrections to the 4-Fermi coupling. At the same order in $N$ as in (\ref{eq:betaBCS}) there is a geometric series of fermion bubbles, as well as anomalous dimension insertions in internal fermion lines. These are automatically resummed into the solution of the RG $\beta$ function. Subleading in $N$ effects come from vertex corrections (as before) as well as higher loop contributions containing BCS interactions that are not sums of fermion bubbles. This large $N$ suppression is a consequence of the non-planarity of the BCS coupling in our theory. We then conclude that (\ref{eq:betaBCS}) is exact at large $N$. The last aspect to understand is whether the 4-Fermi interaction corrects the local non-Fermi liquid behavior of Sec.~\ref{subsec:localNFL}.\footnote{For instance, the ``sunrise'' diagram with two 4-Fermi insertions contributes to the fermion two-point.}
 All such contributions are again subleading at large $N$, again due to the non-planarity of the BCS interaction (recall that the non-Fermi liquid behavior above arises at the planar level). 
 
Another way of organizing these quantum corrections is to study the Schwinger-Dyson equation for the superconducting gap together with (\ref{SD}). This approach will be presented in\cite{RTWlong}, with conclusions that are consistent with the present renormalization group results.
  
%%%%%%%%%%%%%%%%%%%%%%%%%
%%%%%%%%%%%%%%%%%%%%%%%%%
%%%%%%%%%%%%%%%%%%%%%%%%%
%%%%%%%%%%%%%%%%%%%%%%%%%
\section{Quantum criticality and fate of superconductivity}\label{sec:SC}

In this section, we consider the consequences of Eq.~(\ref{eq:betaBCS}). We start by describing, at a heuristic level, the various possible fates of the BCS coupling encoded in this equation.  We then perform a more detailed RG study and discuss the phase diagram of the theory.

%%%%%%%%%%%%%%%%%%%%%%%%%
%%%%%%%%%%%%%%%%%%%%%%%%%
\subsection{Qualitative analysis of the BCS interaction}\label{subsec:qualitative}

For the present analysis, we treat $\gamma$ as an independent parameter to exhibit more clearly the effects from anomalous dimension corrections (in our theory, $\gamma=\alpha/2$). Furthermore, at this heuristic level we will ignore effects from the running of $\alpha$; these will be incorporated below in a more detailed RG treatment.

First, for simplicity consider the case of $\alpha  = \gamma = 0$.  In this case, the scalar is effectively absent from the theory and we recover the marginally relevant flows of the BCS coupling in a Fermi liquid.  Attractive interactions grow under the RG whereas repulsive interactions weaken.  The only difference here, is that BCS couplings are $N$ suppressed  because of the large $N$ limit we have taken.  Thus, there is a limiting case where ordinary Fermi liquid theory is captured.  

Consider next the regime where the term proportional to $\gamma \lambda$ is subdominant.  In this case,  the effect of the anomalous dimension can be neglected but the term $\propto \alpha$  alters the qualitative nature of the superconducting instability. There is an exponentially enhanced pairing instability, and the pairing scale far exceeds the scale at which the Landau quasiparticles would have been destroyed. The inverse correlation length at which $\lambda \gg 1$ in the IR is of order
\be\label{eq:muSon}
\mu_\text{sc} \sim  \exp\left[-\sqrt{\frac{N}{\alpha}} \frac{\pi}{2}\right]\Lambda\,.
\ee
This is what one finds in color superconductivity, and is also what has been recently been reported in Ref. \cite{Metlitski} in the context of 2d quantum criticality. 
Next, consider the regime in which the anomalous dimension plays the most important role.  In this case, one can neglect the higher order terms of the $\beta$ function, and one finds a stable non-Fermi liquid metal with {\it zero} BCS couplings.  

\begin{figure}[h!]
\centering
\includegraphics[width=0.6\textwidth]{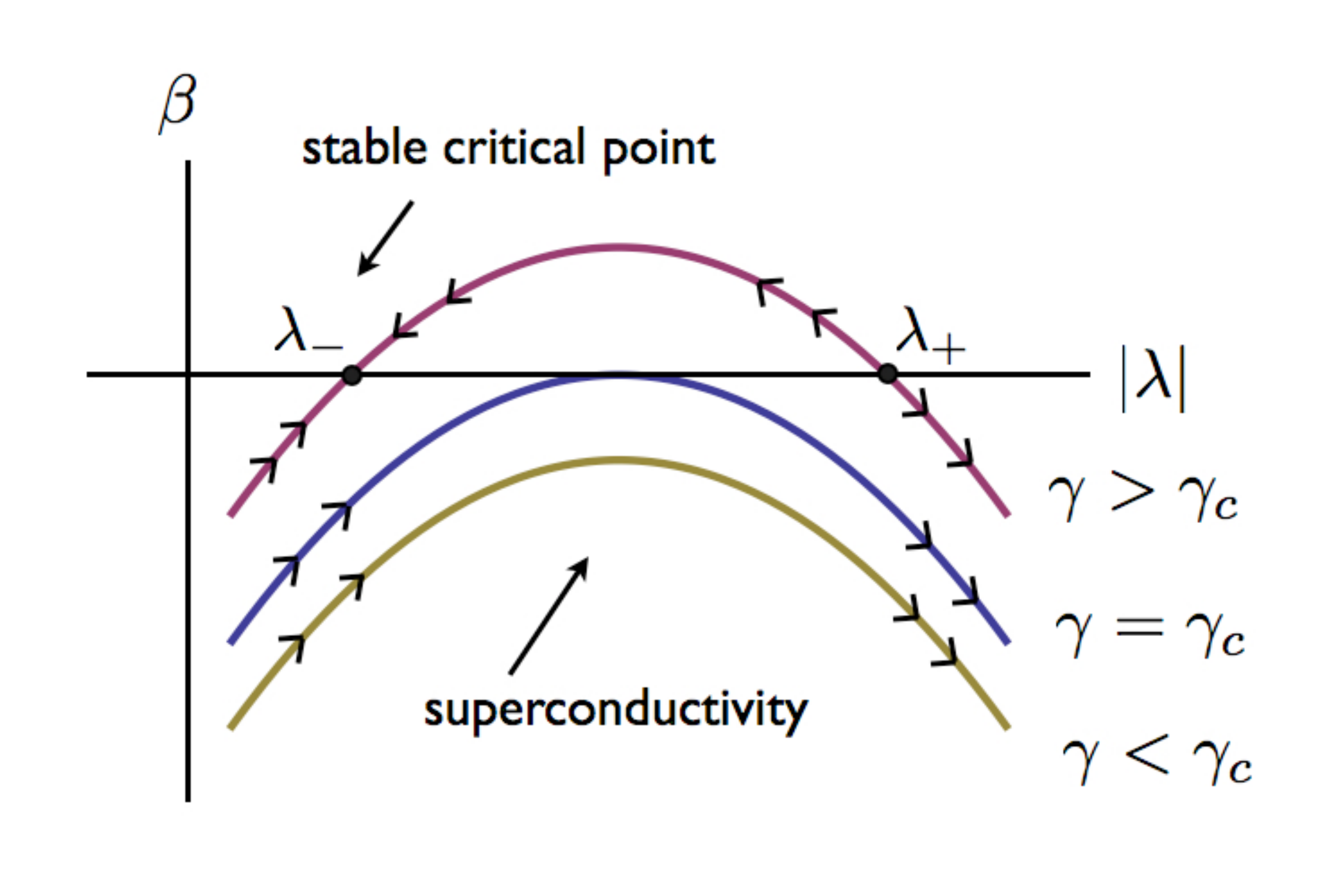}
\caption{\small{BCS $\beta$ function for different values of the anomalous dimension. The two real roots at $\gamma> \gamma_c$ give stable and unstable fixed points. These merge at $\gamma=\gamma_c$ and then annihilate; this gives rise to a superconducting instability for $\gamma<\gamma_c$.}}\label{fig:quadratic-beta}
\end{figure}

Finally, consider the full expression, where all three terms play an important role.  Now, there are competing, and offsetting effects between the anomalous dimension, and the enhanced pairing tendency. 
Since the $\beta$ function is quadratic in $\lambda$, depending on its discriminant there are three different possibilities, illustrated in Fig. \ref{fig:quadratic-beta}.

Neglecting the running of $\alpha$ (to be incorporated shortly), we can write the zeroes as
\begin{equation}
\lambda_{\pm} = \lambda_0 \left( 1 \pm \sqrt{1- \gamma_c^2/\gamma^2} \right)
\end{equation}
where $\lambda_0 = 2 \pi^2 N \gamma$, and $\gamma_c  = \sqrt{\alpha/N}$.  The quantity $\gamma_c$ plays the role of a ``critical anomalous dimension" which separates two regimes. The fixed points are only physical when $\gamma > \gamma_c$, that is, when the zeroes of the $\beta$ function occur on the real axis.  In this case, there is a UV fixed point $\lambda_+$, and an IR fixed point $\lambda_-$. We should emphasize that the critical behavior associated to $\lambda_-$ is qualitatively different from a non-Fermi liquid fixed point that is local on the Fermi surface, since it affects correlation functions with support on antipodal points of the surface. We will discuss further properties of this fixed point in the next section.

As $\gamma \rightarrow \gamma_c$, the two fixed points meet at $\lambda_0$. Finally, when $\gamma < \gamma_c$, the zeroes move off the real axis, and fixed points no longer occur.  Note that near the IR fixed point, $\gamma \sim \epsilon$, whereas $\gamma_c \sim \sqrt{\epsilon/N}$.  Thus, for sufficiently large $N$, the critical value of the anomalous dimension needed to have finite BCS fixed points can be made arbitrarily small.  Therefore, in this theory, $N$ acts as a tuning parameter in the space of theories for $\gamma_c$. 

Next, consider what happens as $\gamma \to \gamma_c$ from above.  In this limit, the UV and IR fixed points approach one another, and when this ratio becomes unity, the fixed points annihilate. Once this happens, for $\gamma< \gamma_c$ the system develops a superconducting instability, and the metallic phase is lost. In this case, the fermion anomalous dimension is not strong enough to avoid a superconducting instability. The inverse correlation length associated with the BCS coupling can be estimated as follows (see also Ref. \cite{Kaplan:2009kr} where such behavior is studied in detail):
\be\label{eq:muBKT}
\xi^{-1}_\text{sc} \simeq \Lambda \exp{\left[ \int_{\lambda_{uv}}^{\lambda_{ir}} \frac{d \lambda}{\beta(\lambda, \alpha)}  \right]} \simeq \exp\left[-\frac{\pi}{2\sqrt{\gamma_c^2-\gamma^2}} \right]\Lambda\,.
\ee
The correlation length diverges exponentially as $\gamma_c/\gamma \rightarrow 1$, signaling a continuous phase transition that separates the quantum critical and superconducting states.  This behavior is similar to the way in which the correlation length of the 2d XY model diverges at the BKT transition.  The analog of critical temperature is played here by $\gamma_c$.

Notice that for $\gamma \to 0$, (\ref{eq:muBKT}) reproduces the boson-enhanced scale (\ref{eq:muSon}). On the other hand, as $\gamma \sim \gamma_c$, we find strong non-Fermi liquid corrections to the superconducting order parameter, which eventually destroy it via a continuous phase transition. Passing this phase transition obtains a quantum critical point characterized by non-Fermi liquid exponents for the quasiparticle dimension, Yukawa coupling, and BCS interaction. This QCP is characterized by a finite BCS coupling, with no superconductivity, and with a power-law behavior for the superconducting correlation function.

%%%%%%%%%%%%%%%%%%%%%%%%%
%%%%%%%%%%%%%%%%%%%%%%%%%
\subsection{Quantum criticality and superconductivity}\label{subsec:qcp-sc}

Before, we presented a qualitative discussion of Eq.~(\ref{eq:betaBCS}). Our task now will be to study in more detail the phase structure and low energy dynamics as a function of $N$ and $\epsilon$

To begin with, let us start from the non-Fermi liquid fixed point at $\alpha=2\gamma=\epsilon/3$. Then, the discriminant of the BCS $\beta$ function vanishes at
\be
\epsilon N =12\,.
\ee
For $N > 12/\epsilon$, the BCS coupling flows to the stable IR fixed point
\be
\lambda_{-}=\frac{\pi^2}{3} \epsilon N (1-\sqrt{1-12/(\epsilon N)})\,.
\ee
Note that $\lambda_- \sim O(1)$ over all the critical range. An important point here is that the attractive fixed point has a finite domain of attraction: for sufficiently large initial values of $\lambda$, the suppression from the anomalous dimension term does not set in fast enough and $\lambda$ will diverge in the IR, signaling a superconducting instability. This is due to the existence of the unstable fixed point at $\lambda_+$. The size of the domain of attraction at the NFL fixed point is of order $|\lambda_+ - \lambda_-|$. This is much bigger than 1 for $N \gg 12/\epsilon$, and shrinks to zero as the continuous transition is approached at $N=12/\epsilon$. These results are explained in more detail in the Appendix.

It is possible to solve exactly the coupled system of equations
\bea
\mu\frac{d\alpha}{d\mu}&=&-\frac{\epsilon}{3}\alpha+\alpha^2\nonumber\\
\mu\frac{d\lambda}{d\mu} &=& -2 \pi^2 \alpha+\alpha \lambda - \frac{\lambda^2}{2\pi^2 N}
\eea
in terms of hypergeometric functions. The boundary conditions for the RG are imposed at some high scale $\mu=M$ which should be below $M_D$ given our approximations on the $z_b=3$ boson scaling. In accordance with our previous analysis, this obtains a fixed point $\alpha=\epsilon/3$ and $ \lambda=\lambda_-$ for $N>12/\epsilon$, a superconducting phase for $N<12/\epsilon$, and a continuous phase transition at $N=12/\epsilon$. 

Near the continuous phase transition at $N \epsilon =12$, the scale of the superconducting instability (namely the scale at which $\lambda$ diverges)  is found to be
\be\label{eq:muBKT2}
\mu_\text{sc} \simeq  \exp\left[-\frac{\pi}{2}\sqrt{\frac{3N}{\epsilon}}\frac{1}{\sqrt{1-\epsilon N/12}} \right]\Lambda\,.
\ee
For $\epsilon N \ll 12$ this shows the enhancement due to boson exchange; however, as $\epsilon N \to 12$ non-Fermi liquid effects dominate over this enhancement and destroy the superconducting parameter by a characteristic BKT scaling. This is one of our main results regarding the competition between superconducting and non-Fermi liquid effects due to a critical boson.

\begin{figure}[h!]
\centering
\includegraphics[width=0.4\textwidth]{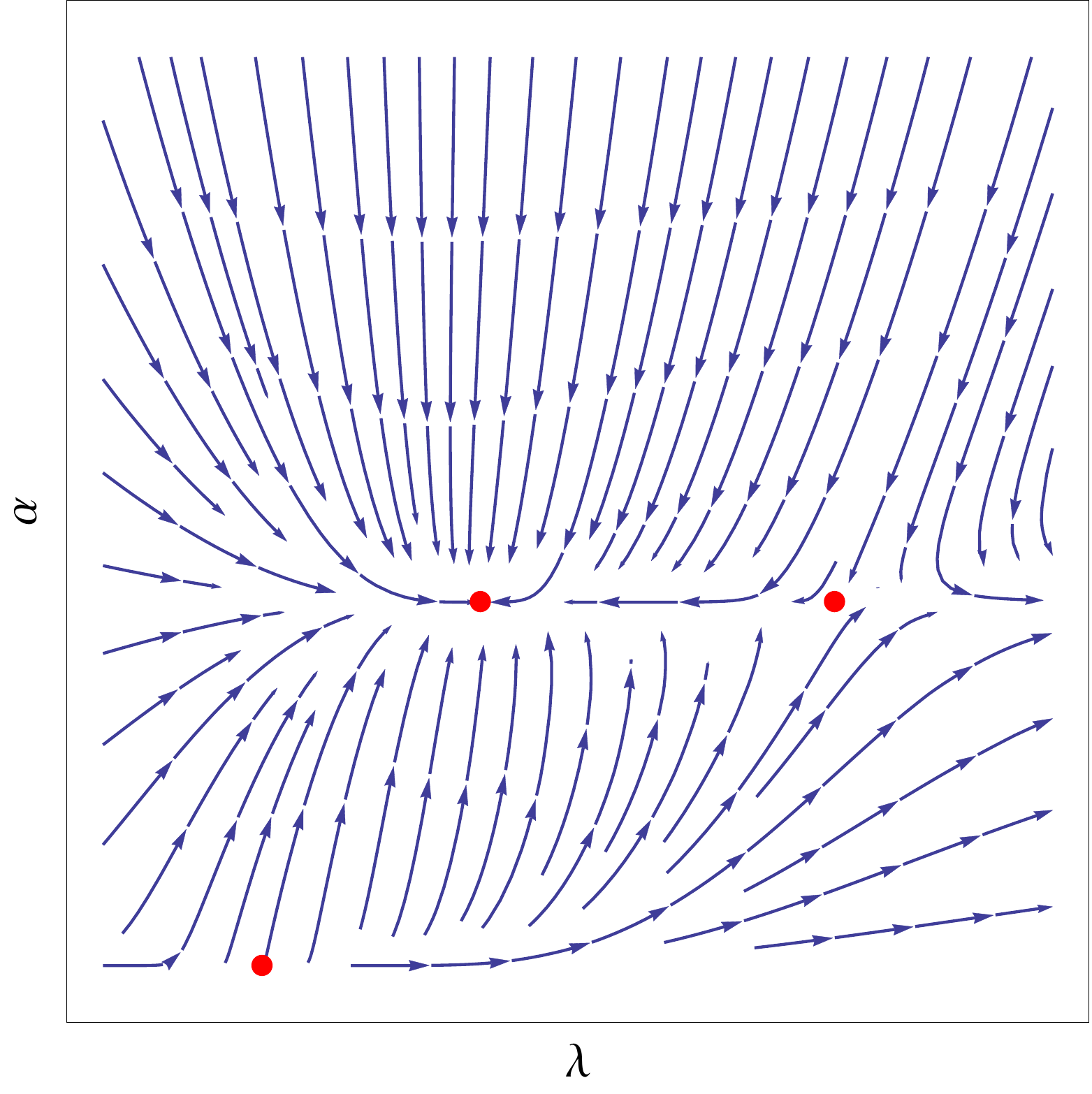}
\includegraphics[width=0.4\textwidth]{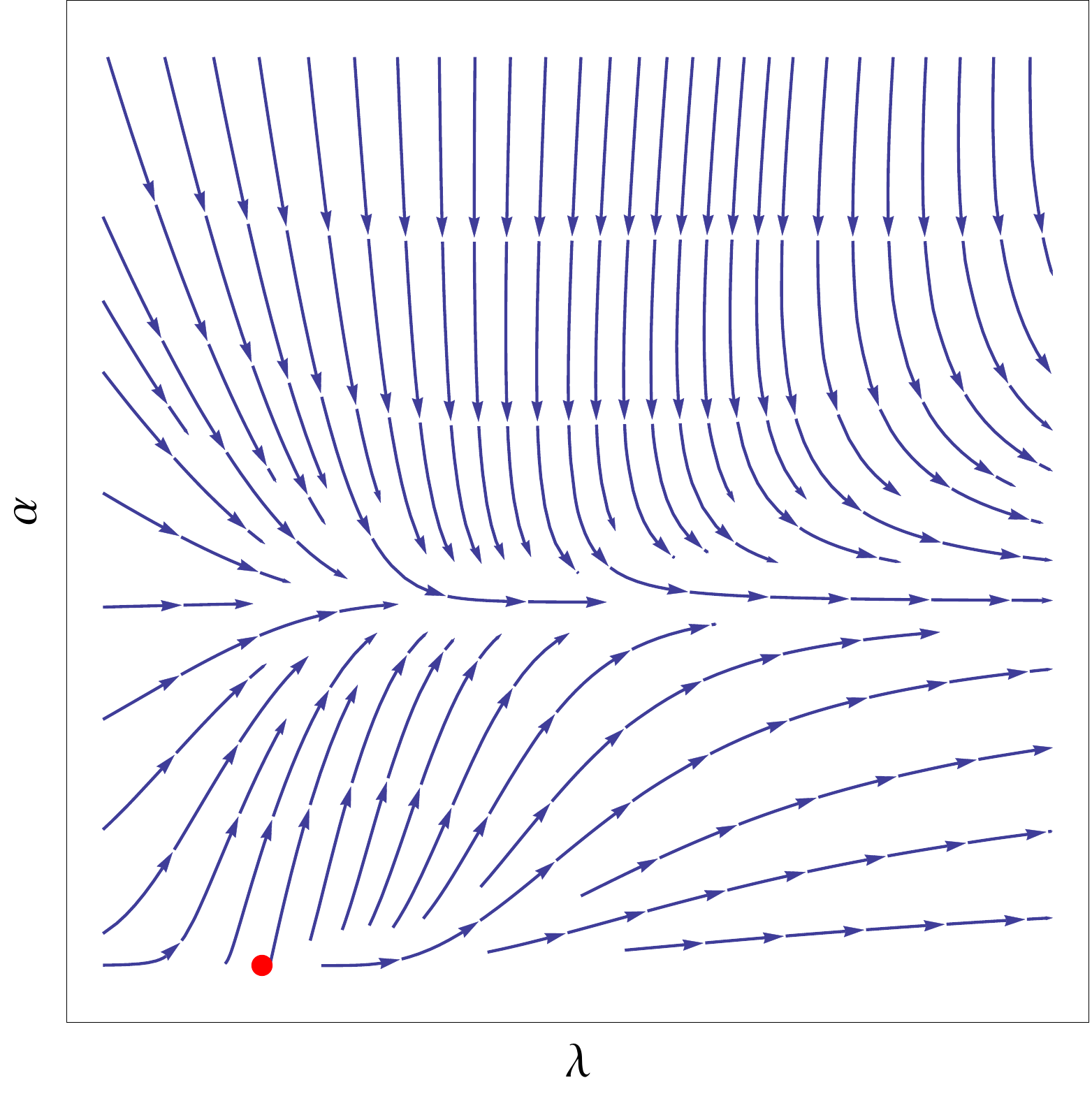}
\caption{\small{Streamlines for $(\beta_\lambda, \beta_\alpha)$, with $N>12/\epsilon$ (left), and $N<12/\epsilon$ (right). The red dots are fixed points; we show the stable, unstable and Gaussian fixed points in the left panel, and the Gaussian fixed point in the right panel. We thank A. Maharaj for help in generating this plot.}}\label{fig:2dRG}
\end{figure}

An illustrative way of presenting these RG results is in terms of streamlines for $(\beta_\lambda, \beta_\alpha)$ as a function of $(\lambda, \alpha)$. The left panel in Fig.~\ref{fig:2dRG} shows the case  $N>12/\epsilon$, and the red points correspond to the stable, unstable and Gaussian fixed points. The flows in the superconducting range $N<12/\epsilon$ are presented in the right panel, where we also show the Gaussian fixed point.

%%%%%%%%%%%%%%%%%%%%%%%%%
%%%%%%%%%%%%%%%%%%%%%%%%%
%%%%%%%%%%%%%%%%%%%%%%%%%
\section{Discussion}\label{sec:discussion}

In this paper, we have  studied a class of quantum metals (such as the Ising nematic quantum phase transition), obtained by coupling a Fermi surface to a nearly critical bosonic order parameter which preserves translation invariance, and condenses at zero momentum. We analyzed the interplay between superconductivity and non-Fermi liquid effects in a theoretically controlled setup depending on $N$ (the rank of an internal global symmetry) and $\epsilon$ (the deviation from spatial dimension $d=3$). We found a novel class of fixed points, stable against the superconducting instability, where the BCS interactions flow to scale invariant values. These QCPs display non-Fermi liquid behavior in observables that are local on the Fermi surface (the anomalous dimension and Fermi velocity) but also in operators that combine fermions on antipodal points, such as the Cooper pair field or the BCS operator. We also showed that for sufficiently small $N$ a superconducting instability sets in, via a continuous phase transition.
We next consider the possible relevance of our findings to the experimental observations of superconducting domes near quantum critical points in a broad class of correlated electron materials.  

\begin{figure}[h!]
\centering
\includegraphics[width=0.9\textwidth]{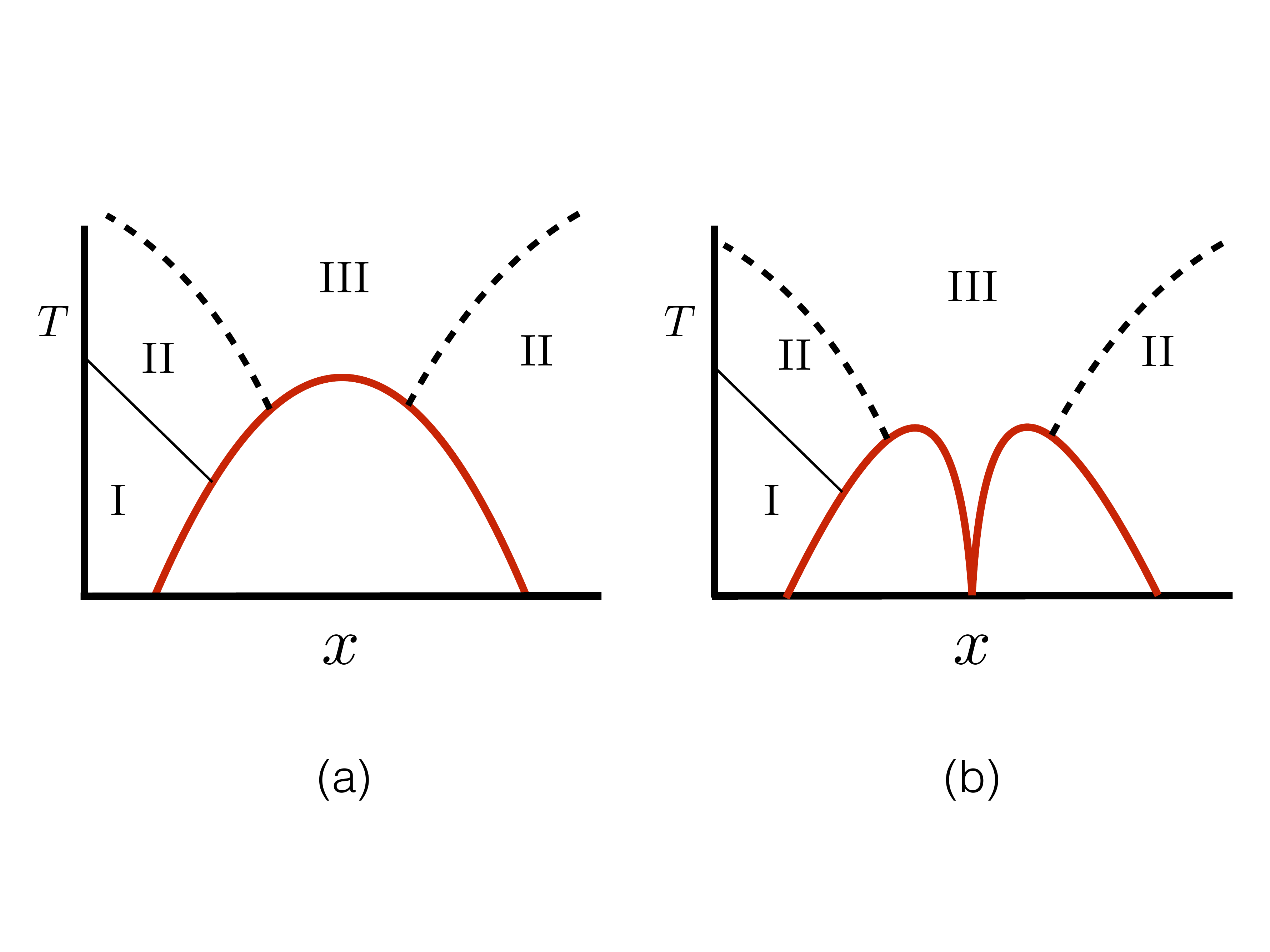}
\caption{\small{Fate of the superconducting domes for different values of $N \epsilon$: (a) $N \epsilon \ll 1$ (left),  and (b) $N \epsilon >12$.  Here the various labels denote the following: the ordered phase (I), Fermi liquid (II), and non-Fermi liquid (III).}}\label{fig:domes}
\end{figure}

We consider the phase diagram as a function of temperature and the parameter that tunes the boson to criticality (e.g. doping, pressure), which we label $x$. For $N \ll 1/\epsilon$, the massless boson produces a strong enhancement in the BCS interaction, but a negligible non-Fermi liquid anomalous dimension. In this case we find a superconducting dome covering the critical point. As $N$ is increased, non-Fermi liquid effects become stronger, with the result that the fermion anomalous dimension tends to make the 4-Fermi attraction irrelevant and decreases the scale of superconductivity. Finally, for $N > 12/\epsilon$, superconductivity is destroyed and the non-Fermi liquid fixed point emerges. The competition between non-Fermi liquid and superconducting fluctuations is summarized in the scaling (\ref{eq:muBKT2}) for the gap. These different regimes are presented schematically in Fig. \ref{fig:domes}.

In this work, we have considered a large $N$ theory where the fermion is in the fundamental of $SU(N)$ and the scalar is in the adjoint. The $N \times N$ bosonic degrees of freedom were crucial for obtaining a quantum critical point stable against the BCS interaction. Another possibility, discussed before in e.g.\cite{Altshuler1994, Polchinski1994, Nayak1994, Nayak1994a, Lee2009, Mross2010, Metlitski}, is to introduce a large number $N_F$ of fermionic fields while the order parameter remains a singlet. In such theories, it seems unlikely to obtain superconducting domes that condense out of a non-Fermi liquid normal state.  The primary reason for this is that the fermion anomalous dimension in such theories is proportional to $\alpha/N_F$.  By contrast, in our formulation,   the fermion anomalous dimension $\gamma \sim \alpha$.  Comparing scales, in large $N_F$ theories, 
\be
\mu_{NFL} \sim e^{-\frac{N_F}{\alpha}}\Lambda
\ee
whereas the scale at which the superconducting instability develops is the same as in our theory,
\be
\mu_\text{sc} \sim e^{-\frac{\pi}{2}\sqrt{\frac{N_F}{\alpha}} }  \Lambda\,.
\ee
It therefore is virtually impossible in a perturbative framework for non-Fermi liquid behavior to occur at scales above the superconducting instabilities in the large N limit of this class of theories.  A similar conclusion, though from a somewhat different approach, has been reported in\cite{Metlitski}.  

In future work, we wish to study the phenomenological consequences of the IR stable fixed point with finite BCS coupling.  In particular, it would be interesting to determine the effect of a scale-invariant BCS interaction on thermodynamics (e.g. heat capacity) and transport (e.g. magnetoresistance, resistivity) properties.  We wish also to study the effect of a magnetic field in our scenario.  For instance, in the regime of $N,\epsilon$ where there is a superconducting dome enveloping a quantum critical point, it is natural to ask what the properties of the system are when a magnetic field is used to destroy superconductivity. It is conceivable that when the superconducting dome is destroyed by a magnetic field, there still remain substantial superconducting fluctuations governed by the BCS coupling; we can then ask whether the Cooper pair fields retain power-law correlations. 

We also found regimes where even though $N \epsilon > 12$, for sufficiently small initial $\alpha$ or for large enough $\lambda$, the system still flows towards the superconducting phase; see Fig. \ref{fig:2dRG}. This gives to a superconducting instability deep inside the non-Fermi liquid state. It would be interesting if this is of relevance to the phase diagram of some of the high $T_c$ superconductors. Finally, the analysis of the superconducting phase using the gap equation, and its connection with the present RG approach, will be discussed in\cite{RTWlong}.

%%%%%%%%%%%%%%%%%%%%%%%%%
%%%%%%%%%%%%%%%%%%%%%%%%%
%%%%%%%%%%%%%%%%%%%%%%%%%

\acknowledgments{We thank  A. Chubukov, 
L. Fitzpatrick,
H. Goldman,
S. Kachru,
S. S. Lee, and
A. Maharaj,
for interesting discussions and comments on the manuscript, and especially A. Chubukov for ongoing discussions on superconductivity and non-Fermi liquids.
SR is supported by the DOE Office of Basic Energy Sciences, contract DE-AC02-76SF00515, a SLAC LDRD grant on `non-Fermi liquids', the John Templeton Foundation, and the Alfred P. Sloan Foundation. GT is supported by CONICET, and PIP grant 11220110100752. HW is supported by a Stanford graduate fellowship. SR acknowledges inspiring conversations with M. Night.}

\appendix

%%%%%%%%%%%%%%%%%%%%%%%%%
%%%%%%%%%%%%%%%%%%%%%%%%%
%%%%%%%%%%%%%%%%%%%%%%%%%
\section{Scaling analysis}\label{app:scaling}

In this Appendix we construct the scaling theory (within the spherical RG) that agrees with the form of the SD quantum corrections.

In the presence of a nontrivial dynamical exponent $z_b$, the RG approach of Ref.~[\onlinecite{Fitzpatrick:2014cfa}] needs to be modified. First, the scaling of the fermions is not changed by $z_b$: from the action near the Fermi surface,
\be
S_f =- \int dp_0 dp_\perp d^{d-1}\hat n\, \psi^\dag(ip_0-vp_\perp) \psi\,,
\ee
we read off the scaling dimensions
\be
[p_0]= [p_\perp]=1\;,\;[\psi(p)]=-3/2\,.
\ee

On the other hand, the bosonic momenta that dominate quantum corrections appear as differences of close-by fermionic momenta, as can be seen from the cubic interaction,
\be
S_\text{Yuk}= \int dp_0 dp_0'\, d^dp\, d^dp' \, g\, \phi(p'-p) \psi^\dag(p') \psi(p)\,,
\ee
Given $\vec p = \hat n (k_F +p_\perp)$, let us decompose the other fermion momentum as
\be
\vec p\,' = \hat n' (k_F +p_\perp') \approx \hat n (k_F +p_\perp')+k_F\delta \hat n \,.
\ee
The boson momentum decomposed with respect to the local Fermi surface direction $\hat n$ then satisfies
\be
\vec q = \hat n q_\perp +\vec q_\parallel\;,\; q_\perp =p_\perp-p_\perp'\;,\; \vec q_\parallel= k_F \delta \hat n\,.
\ee
Therefore, $[q_0]=[q_\perp]=1$, and it remains to understand how to scale $\delta \hat n$.

The scaling of $\vec q_\parallel$ is determined by the $z_b=3$ boson propagator (\ref{eq:Dz3}). Since $[q_\perp]=1$, it is $q_\parallel$ that is affected by the dynamical exponent, and hence is the component that dominates the momentum transfer. We conclude that for bosonic momenta,
\be\label{eq:boson-scalingapp}
[q_0]=[q_\perp]=1\;,\;[q_\parallel]=1/3\;,\;[\phi(q)]=-\frac{10-\epsilon}{6}
\ee
for $d=3-\epsilon$. The scaling of the bosonic momenta has become anisotropic due to the dynamical exponent.

From a purely bosonic point of view it seems somewhat artificial to select a direction $\hat n$ and scale the two components $q_\perp$ and $q_\parallel$ differently, as this breaks the isotropy of the boson dispersion relation. Our statement, however, is that this is the scaling that will dominate inside correlation functions, where the boson momentum behaves as a difference of two fermionic momenta. In many condensed matter systems, the bosons indeed represent fermionic collective modes, hence it is very natural to identify their momenta with the difference between those of fermions. The scaling of $q_\parallel$ is then equivalent to scaling \textit{differences} in fermionic angles $k_F\delta \hat n$, a process that determines the size of the Fermi surface patch that couples more relevantly to a given fermion $\psi(p,\hat n)$. We note the related RG analysis of the fermion-boson system in the patch picture~\cite{Yamamoto2010}, though we stress that the spherical RG being used here is not the same as the patch scaling of \cite{Polchinski1994, Nayak1994, Nayak1994a}.

With these scalings, the classical dimension of $g$ calculated from
\be
S_b= \int dp_0 dp_0' dp_\perp dp_\perp'\, d^{2-\epsilon}(\hat n + \hat n')d^{2-\epsilon}(\hat n - \hat n')\,\, g\, \phi(p'-p) \psi^\dag(p') \psi(p)\,.
\ee
becomes
\be\label{eq:gscalingapp}
[g]=\frac{\epsilon}{6}\,.
\ee

%%%%%%%%%%%%%%%%%%%%%%%%%
%%%%%%%%%%%%%%%%%%%%%%%%%
%%%%%%%%%%%%%%%%%%%%%%%%%
\section{RG solution for the BCS coupling}\label{app:lambda}

We found that the stable fixed point $\lambda_-$ has a finite domain of attraction due to the existence of the unstable fixed point at $\lambda_+$. One interesting consequence of this is that we could have an RG trajectory that ends in a superconducting instability even if $N > 12/\epsilon$. This may then realize a superconducting dome condensing out of a non-Fermi liquid and could be of relevance for certain strongly correlated materials. For this reason, in this Appendix we discuss in more detail how this occurs in a simple case.

By fixing $\alpha$ to its critical value $\alpha=\epsilon/3$, we can solve the RG analytically across the transition, starting from arbitrary UV boundary condition $\lambda(M)=\lambda_0$. Denoting $\mu=M e^{-t}$, the solution takes the form:
\bea
\lambda(t)&=& \pi^2\alpha N\left\lbrace1-\sqrt{\alpha N-4}\tanh{\left(t\sqrt{\frac{\alpha^2}{4}-\frac{\alpha}{N}}+\frac{1}{2}\log{\left(\frac{\lambda_+-\lambda_0}{\lambda_0-\lambda_-}\right)}\right)}\right\rbrace\,\; \alpha N>4,\;\lambda_-<\lambda_0<\lambda_+ \text \nonumber\\  
&=&\pi^2\alpha N\left\lbrace1-\sqrt{\alpha N-4}\coth{\left(t\sqrt{\frac{\alpha^2}{4}-\frac{\alpha}{N}}+\frac{1}{2}\log{\left(\frac{\lambda_+-\lambda_0}{\lambda_--\lambda_0}\right)}\right)}\right\rbrace\,\; \alpha N>4, \text{ otherwise }\nonumber\\ 
 &=&\pi^2\alpha N\left\lbrace1-\sqrt{4-\alpha N}\cot{\left(t\sqrt{\frac{\alpha}{N}-\frac{\alpha^2}{4}}+\tan^{-1}\left(\frac{\sqrt{\frac{4}{\alpha N}-1}}{1-\frac{\lambda_0}{\pi^2\alpha N}}\right)\right)}\right\rbrace\,\;\alpha N < 4\,.
\eea

From these solutions we can see that for $\alpha N>4$, the only possible pole occurs when the argument of the $\coth$ function becomes zero. Since the log term in the argument is positive for $\lambda_0<\lambda_-$, for $t>0$ a pole can only exist for $\lambda_0>\lambda_+$. The corresponding superconducting scale is at: 
\be
\mu^+_{sc}= M \left(\frac{\lambda_0-\lambda_+}{\lambda_0-\lambda_-}\right)^{\frac{1}{\sqrt{\alpha^2-\frac{4\alpha}{N}}}}
\ee
Viewing $\lambda_0$ as a tuning parameter, there is a phase transition at $\lambda_c=\lambda_+$, whose order is given by $\frac{1}{\sqrt{\alpha^2-\frac{4\alpha}{N}}}$. The order diverges as we tune $\alpha N$ towards the critical value $4$. For $\alpha N <4$, the superconducting scale is given by:
\be
\mu^-_{sc}=M \exp{\left[-\frac{2\pi}{\sqrt{\frac{4\alpha}{N}-\alpha^2}}\right]} f(\lambda_0)\,,
\ee
where
\be
f(\lambda_0)= \exp{\left[-\tan^{-1}{\left(
\frac{\pi^2\sqrt{4\alpha N-\alpha^2 N^2}}{\pi^2\alpha N-\lambda_0}\right)}\right]}^{1/\sqrt{\frac{\alpha}{N}-\frac{\alpha^2}{4}}}\,.
\ee
We see that once we cross below $\alpha N = 4$, the phase transition in $\lambda_0$ reduces to a step discontinuity across $\lambda_{disc}=\pi^2\alpha N$: there is no choice of $\lambda_0$ that can kill the superconducting instabilities.  

\bibliography{shamit}

\end{document}